\documentclass[aps,pre,twocolumn,superscriptaddress,amsmath]{revtex4}
	\usepackage{bm}
	\usepackage{graphicx}
\newcommand\beq{\begin{equation}}
\newcommand\eeq{\end{equation}}
\newcommand\beqa{\begin{eqnarray}}
\newcommand\eeqa{\end{eqnarray}}

\def\one{{(1)}}
\def\two{{(2)}}

\def\k{{(N)}}

\def\s{\langle\sigma\rangle}
\def\ss{\langle\sigma^2\rangle}
\def\sss{\langle\sigma^3\rangle}
\begin{document} 
\title{Equation of state of a multicomponent 
$d$-dimensional hard-sphere
fluid\\
\small{[Mol. Phys. \textbf{96}, 1--5 (1999)]}}
\author{Andr\'{e}s Santos}
\author{Santos Bravo Yuste}
\affiliation{Departamento de F\'{\i}sica, Universidad de 
Extremadura, 
E-06071 Badajoz, Spain}
\author{Mariano L\'opez de Haro}
\affiliation{Centro de Investigaci\'{o}n en Energ\'{\i }a, 
U.N.A.M.,
Apartado Postal 34, Temixco, Mor.\ 62580, Mexico}
\begin{abstract}
A simple recipe to derive the compressibility factor of a multicomponent 
mixture of $d$-dimensional additive hard spheres in terms of that of the 
one-component system is proposed. The recipe is based (i) on an exact 
condition that has to be satisfied in the special limit where one of the 
components corresponds to point particles; and (ii) on the form of the 
radial distribution functions at contact as obtained from the Percus-Yevick 
equation in the three-dimensional system.
The proposal is examined for hard discs and hard spheres by comparison with 
well-known equations of state for these systems and with simulation data.
In the special case of  $d=3$, our extension 
to mixtures of the Carnahan-Starling equation of state yields a better 
agreement with simulation than the already accurate 
Boubl\'{\i}k-Mansoori-Carnahan-Starling-Leland equation of state.
\end{abstract}

\maketitle

Due to their importance in liquid state theory, for years researchers have 
proposed empirical or semi-empirical
(analytical) equations of state of various degrees of complexity for
one-component hard-sphere fluids. Notable among these, is the celebrated
Carnahan-Starling (CS) equation of state \cite{CS},
which is not only rather simple but also accurate in
comparison with computer simulation data. In the case of hard-sphere
mixtures, the proposals, also empirical or semi-empirical in nature, are
much more limited, with the Boubl\'{\i}k-Mansoori-Carnahan-Starling-Leland 
(BMCSL) equation
 \cite{Bou70} standing out as the usual 
favorite. The situation
for one-component hard-disc fluids is rather similar. Here, no analog of the
CS equation using the 
$\frac{1}{3}\mbox{(v)}+\frac{2}{3}\mbox{(c)}$ recipe has
been derived, due to the absence of an analytical solution of the
Percus-Yevick (PY) equation in this instance. Nevertheless, accurate and 
simple
equations of state have been proposed, such as the popular Henderson
equation \cite{DH} and the recent one by the present authors \cite{SHY}. 
Hard-disc
mixtures, on the other hand, have received much less attention and
the proposed equations of state for these systems are rather scarce. Given
this scenario, the major aim of this paper is to show that, on the 
basis of  a simple recipe, accurate
equations of state of a multicomponent $d$-dimensional additive hard-sphere 
mixture
may be derived, requiring the equation of state of the one-component
system as the only input. The recipe makes use of a consistency condition
that arises in the case that one of the components in the mixture has a
vanishing size, as well as from some insight gained from the form of the
radial distribution functions at contact given by the solution of the 
Percus-Yevick equation for a hard-sphere fluid in three
dimensions \cite{L64}.


Let us consider an $N$-component system of hard spheres in $d$ dimensions.
The total number density is $\rho $, the set of molar fractions is 
$\{x_1,\ldots ,x_N\}$, and the set of diameters is 
$\{\sigma _1,\ldots ,\sigma _N\}$. The volume packing 
fraction
is $\eta =v_d\rho \langle \sigma ^d\rangle $, where  $v_d=(\pi 
/4)^{d/2}/\Gamma
(1+d/2) $ is the volume of a $d$-dimensional sphere of unit diameter
 and $\langle \sigma^n\rangle \equiv \sum_ix_i\sigma _i^n$. 
In the case of a polydisperse mixture ($N\to\infty$) characterized by a size 
distribution $f(\sigma)$,  $\langle \sigma^n\rangle \equiv \int 
d\sigma\sigma^n f(\sigma)$. Our
goal is to propose a {\em simple\/} equation of state (EOS) for the mixture, 
$Z^{(N)}(\eta)$, consistent with a given EOS for a
one-component system, $Z^{(1)}(\eta )$, where $Z=p/\rho k_BT$ is the usual 
compressibility factor. 
 A  consistency condition appears when one of the species, say 
the $N$-th,
has a vanishing diameter, i.e. $\sigma_N\to 0$.
In that case, 
\begin{equation}
Z^{(N)}(\eta)\to
(1-x_N)Z^{(N-1)}( \eta)
+\frac{x_N}{1-\eta }.  \label{1}
\end{equation}
At a more fundamental level, we will consider the contact values of the
radial distribution functions, $g_{ij}^{(N)}(\sigma _{ij})$, the knowledge 
of which implies that of the EOS through the relation 
\begin{equation}
Z^{(N)}(\eta)=1+2^{d-1}\eta
\sum_{i=1}^N\sum_{j=1}^Nx_ix_j\frac{\sigma _{ij}^d}{\langle \sigma ^d\rangle 
}g_{ij}^{(N)}(\sigma _{ij}).
\label{2}
\end{equation}

Taking as a guide the form of $g_{ij}^{(N)}$ obtained through
the solution of the
PY equation for hard spheres ($d=3$) \cite{L64},
we  propose to approximate
$g_{ij}^{(N)}$ by a linear interpolation between 
$g^{(1)}$ and $(1-\eta)^{-1}$, namely
\begin{equation}
g_{ij}^{(N)}(\sigma _{ij})=\frac 
1{ 1-\eta }+\left[ g^{(1)}(\sigma)-\frac 1{1-\eta }\right] 
\frac{\langle
\sigma ^{d-1}\rangle }{\langle \sigma ^d\rangle }\frac{\sigma _i\sigma _j}{ 
\sigma _{ij}}.  \label{4.11}
\end{equation}
When the above ansatz is inserted into equation (\ref{2}), one gets 
\beqa
Z^{(N)}(\eta)-1 &=&\left[ Z^{(1)}(\eta
)-1\right] 2^{1-d}\Delta _0 \nonumber\\
&&+\frac{\eta}{1-\eta }\left( 1-\Delta _0 
+ \frac{1}{2}\Delta _1\right) ,  
\label{4.1}
\eeqa
where 
\beqa
\Delta _p&=&\frac{\langle \sigma
^{d+p-1}\rangle }{\langle \sigma ^d\rangle ^2}\sum_{n=p}^{d-1}
\frac{(d+p-1)!}{n!(d+p-1-n)!}
 \langle \sigma ^{n-p+1}\rangle \langle \sigma ^{d-n}\rangle ,
 \nonumber\\
&& \qquad \qquad
(p=0,1).  \label{4.2}
\eeqa
This form of the EOS complies with the requirement (\ref{1}).
Note that
equation (\ref{4.1}) expresses $Z^\k(\eta)-1$ as a linear combination of 
$Z^\one(\eta)-1$ and $(1-\eta)^{-1}-1$, but the dependence of the 
coefficients on the size distribution is much more involved than in equation 
(\ref{4.11}). The key outcome of this paper is the EOS 
given by equation (\ref{4.1}), in which the compressibility factor of the 
mixture is obtained from that of the one-component system for 
arbitrary values of the dimensionality $d$ and the number of components $N$.
It is worth noticing that in the one-dimensional case, equation (\ref{4.1})
yields the {\em exact\/} result $Z^{(N)}(\eta)=Z^{(1)}(\eta)$.

As a straightforward application of equation (\ref{4.1}), one can easily get
$B_n^{(N)}=v_d^{n-1}\langle \sigma ^d\rangle ^{n-1}\left[ 2^{1-d}\Delta
_0b_n^{(1)}+1-\Delta _0+\frac 12\Delta _1\right]$,
where
the virial coefficients $B_n^{(N)}$ are
defined by 
$Z^{(N)}(\eta)=1+\sum_{n=2}^\infty B_n^{(N)}\rho ^{n-1}$.  
 and where $b_n^{(1)}$ are the reduced virial coefficients of the 
 one-component
system, i.e. 
$
Z^{(1)}(\eta)=1+\sum_{n=2}^\infty b_n^{(1)}\eta ^{n-1}
$.  


We will now focus on the case of hard discs ($d=2$). 
Equation (\ref{4.1}) then becomes 
\begin{equation}
Z^{(N)}(\eta)=Z^{(1)}(\eta 
)\frac{\langle
\sigma \rangle ^2}{\langle \sigma ^2\rangle }+\frac 1{1-\eta }\left( 
1-\frac{ \langle \sigma \rangle ^2}{\langle \sigma ^2\rangle }\right) .  
\label{5}
\end{equation}
The relationship between $Z^\k(\eta)$ and $Z^\one(\eta)$ as given by 
equation (\ref{5}) rests on a different rationale from that pertaining 
to another simple proposal, namely the 
 Conformal Solution Theory (CST)  \cite{HM86,Baus}. In this latter theory, 
 the EOS reads
$Z^{(N)}_{\text{CST}}(\eta)=Z^{(1)}(\eta 
_{\text{eff} })$ with
$ \eta _{\text{eff}}= \frac 12\left( 1+\langle 
\sigma \rangle
^2/\langle \sigma ^2\rangle \right) \eta$,
but we note that this equation {\em does not\/} comply 
with the
general requirement (\ref{1}).
If the one-component system is assumed to be described by the Scaled 
Particle Theory (SPT)
 \cite{RFL}, our extension to mixtures takes on a particularly simple form: 
\begin{equation}
Z^{(N)}_{\text{SPT}}(\eta)
=\frac{1-\left( 1-\langle
\sigma \rangle ^2/\langle \sigma ^2\rangle \right) \eta }{(1-\eta )^2}. 
  \label{5.1}
\end{equation}
This is precisely the true SPT EOS for mixtures  \cite{Baus}, which is 
indeed rewarding. 
The EOS 
$Z^{(1)}(\eta)=\left[1-2\eta+(2\eta_0-
1)\eta^2/\eta_0^2\right]^{-1}$, where $\eta_0=\sqrt{3}\pi/6$ is the 
crystalline close-packing fraction, has been recently proposed by us 
 \cite{SHY} to describe a one-component system. When this EOS, hereafter 
referred to as the SHY EOS following the nomenclature introduced in reference
 \cite{Fpictures}, is substituted into equation (\ref{5}), we obtain the 
following extension:
\begin{equation}
Z^{(N)}_{\text{eSHY}}(\eta)
=\frac{{\langle \sigma
\rangle ^2}/{\langle \sigma ^2\rangle }}{1-2\eta +(2\eta _0-1)\eta ^2/\eta
_0^2}+\frac 1{1-\eta }\left( 1-\frac{\langle \sigma \rangle ^2}{\langle
\sigma ^2\rangle }\right)  .  \label{5.2}
\end{equation}
The well-known Henderson (H) EOS \cite{DH} can also be extended: 
\begin{equation}
Z^{(N)}_{\text{eH}}(\eta)
=\frac{1-\left( 1-\langle
\sigma \rangle ^2/\langle \sigma ^2\rangle \right) \eta +(b_3^{(1)}-3)\left(
\langle \sigma \rangle ^2/\langle \sigma ^2\rangle \right) \eta ^2}{(1-\eta
)^2},   \label{5.3}
\end{equation}
where $b_3^{(1)}=\frac{16}3-\frac 4\pi \sqrt{3}$.
This equation is
quite similar to the one proposed by Barrat {\em et al.} \cite{Baus}, 
the only difference being that the coefficient of $\eta^2$ in the 
numerator is $b_3^{(N)}-1-2\langle \sigma \rangle ^2/\langle \sigma 
^2\rangle$, where  $b_3^{(N)}$ is
the exact (reduced) third virial coefficient. Nevertheless, since 
$b_3^{(N)}$ is well approximated by $1+(b_3^{(1)}-1)\langle \sigma \rangle 
^2/\langle \sigma ^2\rangle$, both EOS 
 are practically indistinguishable.
 Although we could consider the extensions of  other EOS originally 
proposed for a one-component system of hard discs (for a list of many such 
EOS we refer the reader to references  \cite{SHY,Fpictures}), for the sake 
of simplicity we will restrict our analysis to the SPT, eSHY and eH EOS.

Let us now consider the virial coefficients $B_n^{(N)}$ for hard discs. 
It follows that in this case $
B_n^{(N)}=\left(  \pi/ 4\right) ^{n-1}\langle \sigma ^2\rangle ^{n-1}
\left[1+(b_n^{(1)}-1)\langle \sigma \rangle
^2/\langle \sigma ^2\rangle \right]
$.
This equation  yields the exact second virial
coefficient  \cite{Baus}, the higher coefficients being approximate.
In the particular case of a binary mixture, the
composition-independent coefficients $B_{n_1,n_2}^{(2)}$ are defined
through $B_n^{(2)}=\sum_{n_1=0}^n\frac{n!}{n_1!(n-n_1)!}
B_{n_1,n-n_1}^{(2)}x_1^{n_1}x_2^{n-n_1}
$.
According to  equation (\ref{5}), 
\begin{eqnarray}
B_{n_1,n_2}^{(2)} &=&\left( \frac \pi 4\right) ^{n-1}\sigma
_1^{2(n-1)}\alpha ^{2(n_2-1)}\left[ \frac{n_1}n\alpha ^2+\frac{n_2}n\right. 
+\left( b_n^{(1)}-1\right)
\nonumber   \\
&&\left.\times  \left( \frac{n_1(n_1-1)}{n(n-1)}\alpha 
^2 +\frac{2n_1n_2}{n(n-1)}\alpha +\frac{n_2(n_2-1)}{n(n-1)}\right) \right],
\nonumber\\ 
&&
\label{10}
\end{eqnarray}
where $n=n_1+n_2$ and $\alpha =\sigma _2/\sigma _1$. This form
has the same structure as the interpolation formula suggested by Wheatley 
\cite{W98}.
In fact, he proposes an EOS (henceforth labelled as W) of the form
\beq
\label{10.1}
Z^\two_{\text{W}}(\eta)
=\frac{\sum_{n=0}^7 c_n \eta^n}{(\eta-\eta_0)^2},
 \eeq
where the coefficients $c_n$ are chosen so as to reproduce the first 
eight virial coefficients given by the interpolation formula \cite{W98}.


Now, let us consider the case $d=3$. Equation (\ref{4.1}) then
yields 
\begin{eqnarray}
Z^{(N)}(\eta) 
&=&
=1+\left[ Z^{(1)}(\eta
)-1\right] \frac{\langle \sigma ^2\rangle }{2\langle \sigma ^3\rangle ^2} 
\left( \langle \sigma ^2\rangle ^2+\langle \sigma \rangle \langle \sigma
^3\rangle \right)  
\nonumber   \\&&
+\frac \eta {1-\eta }\left[ 1-\frac{\langle \sigma ^2\rangle }{\langle
\sigma ^3\rangle ^2}\left( 2\langle \sigma ^2\rangle ^2-\langle \sigma
\rangle \langle \sigma ^3\rangle \right) \right] .  \label{4.3}
\end{eqnarray}
Using the CS EOS \cite{CS}, $Z^{(1)}(\eta)=(1+\eta +\eta
^2-\eta ^3)/(1-\eta )^3$, the result may be expressed as
\beq
\label{4.5}
Z^\k_{\text{eCS}}(\eta)=Z^\k_{\text{BMCSL}}(\eta)+
\frac{\eta^3}{(1-\eta)^3}\frac{\ss}{\sss^2}\left(\s\sss-\ss^2\right),
\eeq
where the
BMCSL EOS \cite{Bou70} is
\begin{equation}
Z^{(N)}_{\text{BMCSL}}(\eta)=
\frac 1{1-\eta }+\frac{ 
3\eta }{(1-\eta )^2}\frac{\langle \sigma \rangle \langle \sigma ^2\rangle }{ 
\langle \sigma ^3\rangle }+\frac{\eta ^2(3-\eta)}{(1-\eta )^3}\frac{\langle 
\sigma
^2\rangle ^3}{\langle \sigma ^3\rangle ^2}  .
\label{4.6}
\end{equation}
As another, example, let us consider the Carnahan-Starling-Kolafa (CSK) 
EOS \cite{Ko86}, 
$Z^{(1)}(\eta)=[1+\eta +\eta ^2-2\eta ^3(1+\eta )/3]/(1-\eta )^3$. Its extension 
is 
\beqa
Z_{\text{eCSK}}^{(N)}(\eta)&=&Z_{\text{eCS}}^{(N)}(\eta)
+\frac{\eta ^3(1-2\eta )}{ (1-\eta )^3}\frac{\langle \sigma 
^2\rangle }{6\langle \sigma ^3\rangle ^2}\nonumber\\
&&\times \left( \langle \sigma 
^2\rangle ^2+\langle \sigma \rangle \langle \sigma
^3\rangle \right)  ,  \label{4.11_1}
\eeqa
which does not coincide with Boubl\'{\i}k's extension to
mixtures of the CSK EOS \cite{Bou86}:
\beq
\label{4.6_2}
Z^\k_{\text{BCSK}}(\eta)=Z^\k_{\text{BMCSL}}(\eta)+
\frac{\eta^3(1-2\eta)}{(1-\eta)^3}\frac{\ss^3}{3\sss^2}.
\eeq
Recently, Henderson and Chan (HC) have proposed a modification of the BMCSL
EOS \cite{HSW98,Yau96} for the {\em particular\/} case of a binary mixture 
in which the concentration of the large spheres is exceedingly small, 
starting from an {\em asymmetric\/} prescription for the radial
distribution functions at contact. The resulting EOS,
with $\sigma_1\geq\sigma _2$,
 is
\begin{eqnarray}
Z^\two_{\text{HC}}(\eta)&=&Z^\two_{\text{BMCSL}}(\eta)+\frac{4\eta x_1}{\sss}
\left\{x_1\sigma_1^3\left\{\frac{3\eta }{2(1-\eta )^2}
\right.\right.\nonumber\\
&&
\left(1-\frac{\ss}{\sss}\sigma_1\right)+
\frac{\eta ^2}{2(1-\eta )^3} 
\left[1-\left(\frac{\ss}{\sss}\sigma_1\right)^2\right]
\nonumber\\
& &\left.+
\exp\left[ \frac{3\eta }{2(1-\eta )^2}\left( \frac{\langle \sigma
^2\rangle }{\langle \sigma ^3\rangle }\sigma _1-1\right) \right] -1
\right\}\nonumber\\
&&+\frac{\eta^2x_2}{4(1-\eta)^3}\left(\frac{\ss}{\sss}\sigma_2\right)^2
(\sigma_1-\sigma_2)
\nonumber\\
&&\left.\times
\left[(\sigma_1+\sigma_2)^2-\eta
\sigma_2(\sigma_1^2+\sigma_2^2+\sigma_1\sigma_2)\frac{\ss}{\sss}\right]
\right\}.\nonumber\\
&&
\label{4.21}
\end{eqnarray}

\begin{figure}[tbp]
\includegraphics[width=1.0\columnwidth]
{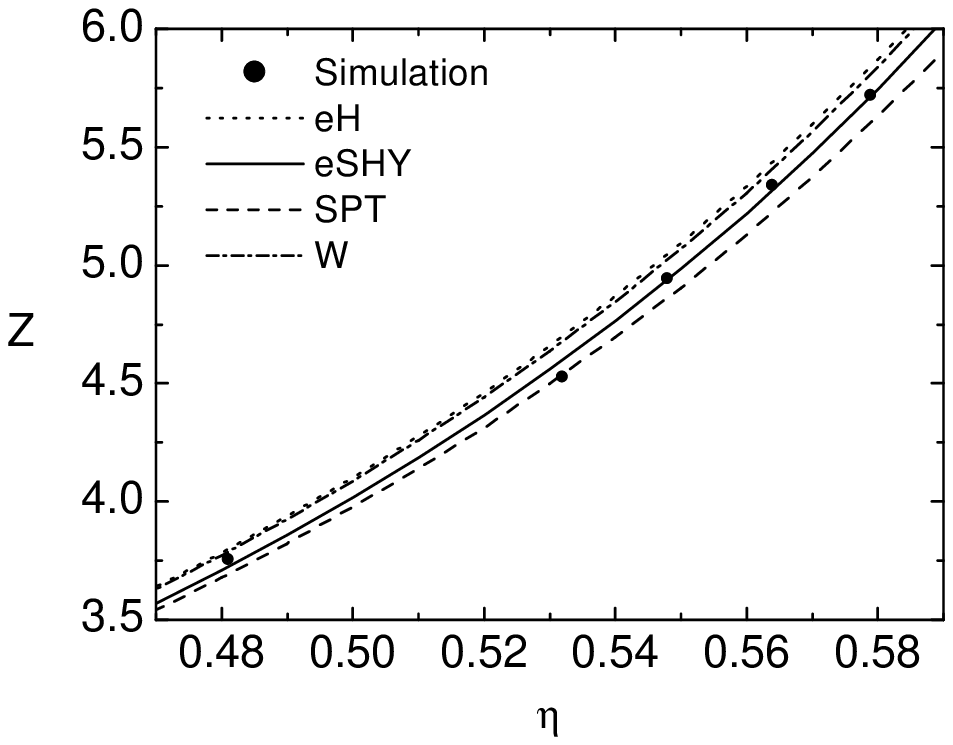}
\caption{Compressibility factor as a function of the packing fraction for  a 
binary mixture of hard discs defined by $x_1=0.351$ and 
$\sigma_2/\sigma_1=0.8$.
\label{fig1}}
\end{figure}
\begin{figure}[tbp]
\includegraphics[width=1.0\columnwidth]
{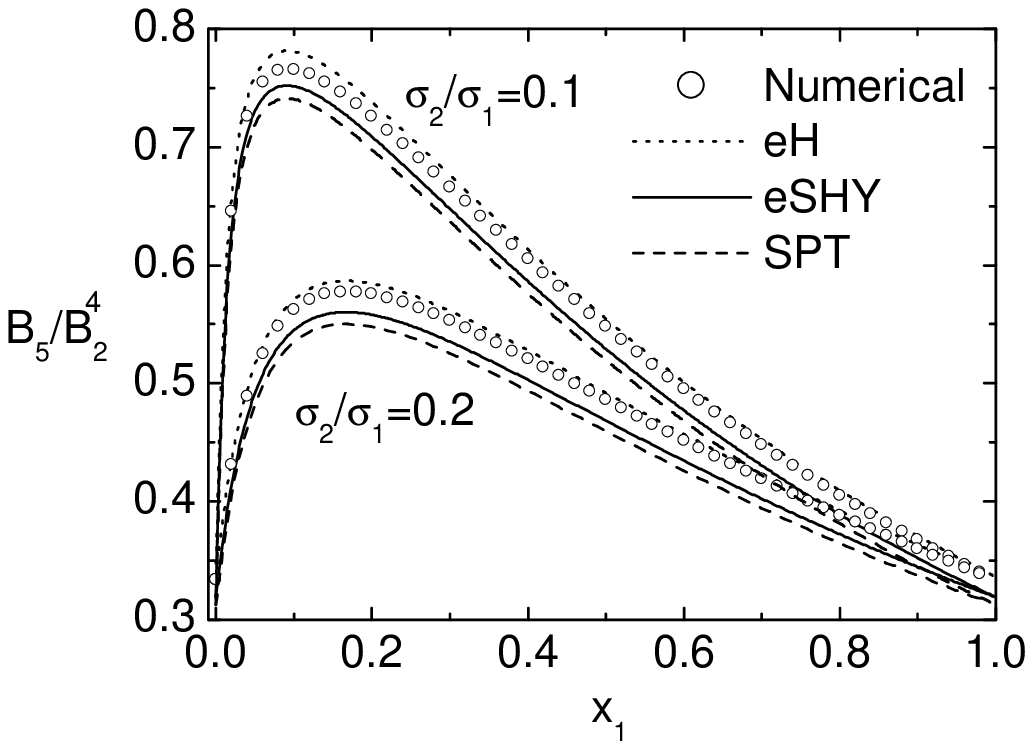}
\caption{Fifth virial coefficient as a function of $x_1$ for two binary 
mixtures of hard discs defined by $\sigma_2/\sigma_1=0.1$ and 
$\sigma_2/\sigma_1=0.2$.
\label{fig3}}
\end{figure}

We shall now perform a
comparison with the (very few) available computer simulation data. We begin
with hard-disc mixtures. In figure \ref{fig1} 
we display the packing-fraction
dependence of the compressibility factor $Z$ for the SPT, W, eH and  eSHY 
EOS, together with the simulation 
results
of Barrat {\em et al.} \cite{Baus}, for the binary mixture defined by $ 
x_1=0.351$ and $\sigma _2/\sigma _1=0.8$. In this case, the
performance of the eSHY EOS is outstanding and clearly superior to all the
other choices. To complete the picture, in figure 
\ref{fig3} we present the
results for the ratio of the fifth virial coefficient to the fourth power of 
the (exact)
second virial coefficient as a function of the larger disc
concentration and for two size ratios. Here, the best agreement with the
numerical data of Wheatley
\cite{WH98} is obtained with the eH EOS, which is not very surprising
since in the one-component case ($x_1=1$) it gives a very good estimate of
this ratio.
 Nevertheless, the overall trends including the position of the maximum
are still captured in all approximations.
\begin{figure}[tbp]
\includegraphics[width=1.0\columnwidth]
{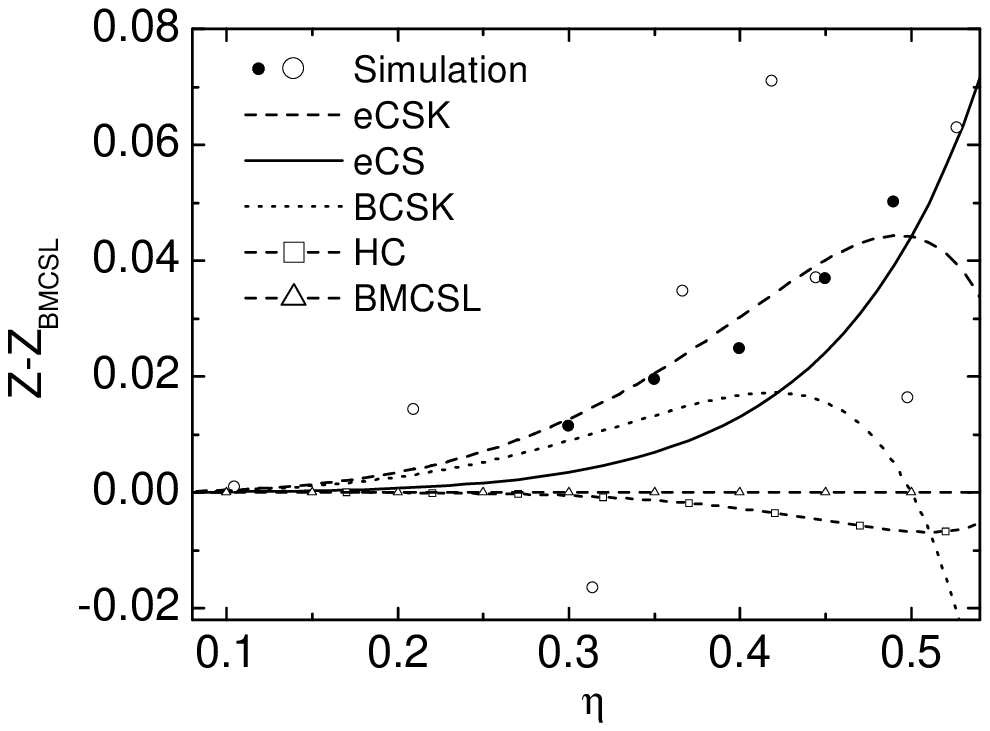}
\caption{Compressibility factor as a function of the packing fraction, 
relative to the BMCSL value, for 
 an equimolar binary mixture of hard spheres with
$\sigma_2/\sigma_1=0.6$.
\label{fig4}}
\end{figure}
\begin{figure}[tbp]
\includegraphics[width=1.0\columnwidth]
{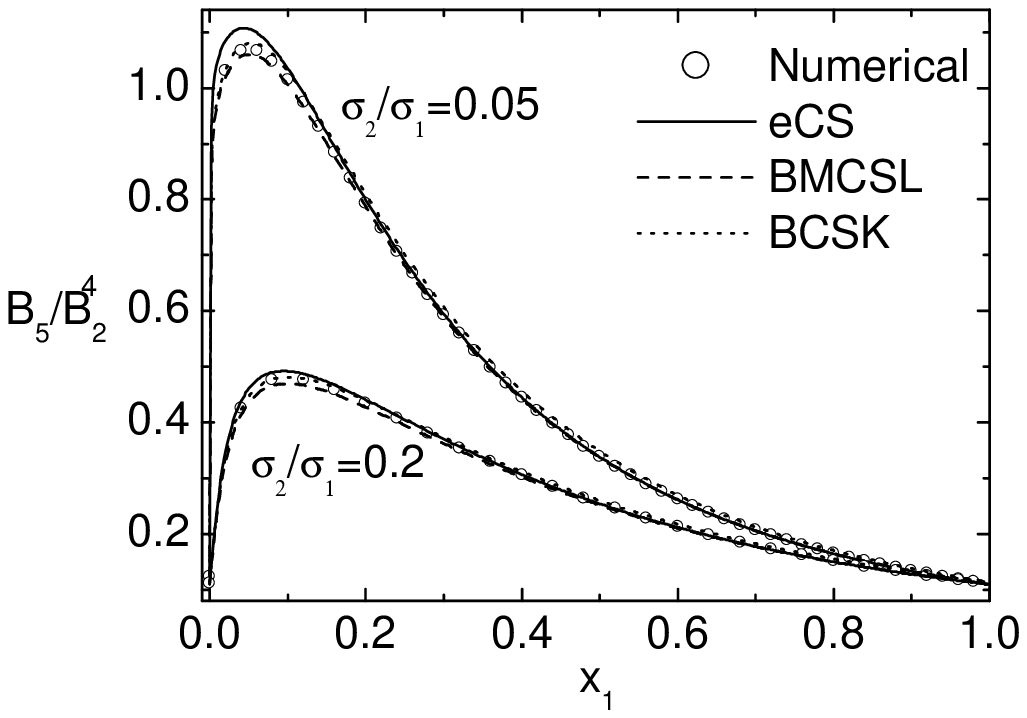}
\caption{Fifth virial coefficient as a function of $x_1$ for two binary 
mixtures of hard spheres defined by $\sigma_2/\sigma_1=0.05$ and 
$\sigma_2/\sigma_1=0.2$.
\label{fig6}}
\end{figure}

As far as hard-sphere mixtures are concerned, the following comments can be
made. 
To our knowledge, only simulation results for binary mixtures have been 
reported.
The most recent data \cite{BMLS96} indicate that the BMCSL EOS 
underestimates the pressure as obtained through simulation.
In fact, the BCSK EOS is geared to correct this deficiency, at least for 
$\eta<0.5$, in a similar fashion as the CSK EOS corrects the CS EOS.
As a general trend, for $\eta<0.5$, our extended EOS, namely the eCS and the 
eCSK, also go in the correct direction.
Moreover, in this density range, 
$Z_{\text{BMCSL}}<Z_{\text{eCS}}<Z_{\text{eCSK}}$ and
$Z_{\text{BMCSL}}<Z_{\text{BCSK}}<Z_{\text{eCSK}}$.
Although limited in scope, the results shown in figure \ref{fig4} illustrate 
these features. Here we have considered an equimolar binary mixture with 
size ratio $\sigma_2/\sigma_1=0.6$.  
As the differences between the values predicted by the various EOS for the
compressibility factor $Z$ are very small,
we have chosen
to present the results, including the simulation data of Yau {\em et 
al.} \cite
{Yau96} (open circles) and Baro\v {s}ov\'{a} {\em et al.} \cite{BMLS96} 
(filled circles), in terms of the
packing fraction dependence of $Z-Z_{\text{BMCSL}}$.
Despite the scatter of the simulation results of Yau {\em et 
 al.} \cite{Yau96}, it is apparent that,
depending on the range, both
the eCSK and eCS EOS seem to do a better job than either the BMCSL, BCSK or 
HC EOS (although in all fairness we should add that the equimolar condition 
is beyond the scope for which the latter EOS was originally devised).
In fact, if one considers a higher density point computed by 
Yau {\em et al.} \cite{Yau96}  
($\eta=0.59$, $Z_{\text{simul}}-Z_{\text{BMCSL}}=0.26$) that is  off 
the scale,  it appears that the overall trend is better captured by the eCS 
EOS, although for $\eta<0.5$ the eCSK should probably be the preferred EOS. 
  The results for the composition dependence of the  fifth virial
coefficient for a binary mixture and two values of $\sigma_2/\sigma_1$
displayed in figure  \ref{fig6} also 
indicate that all the
approximations lead to very good values as compared to the recent numerical
data of Enciso {\em et al.} \cite{Enciso98}, with  a slight
superiority of the BCSK EOS for the region around the maximum. The HC and 
the eCSK results have not been included, since they are almost identical to 
the ones of the BMCSL and the eCS, respectively.

In conclusion, it is fair to state that we have introduced a very simple and
general recipe that allows one to get a reasonably accurate approximation to
the EOS of a multicomponent mixture of $d$-dimensional hard-spheres from any
reasonable EOS of the one-component system. It also seems that, as 
exemplified in
the case of binary three-dimensional hard-sphere mixtures, the more accurate
the EOS of the one-component system,  the better
results the approximation yields. 


Two of us (A. S.) and (S. B. Y.) would like to acknowledge partial support
from the DGES\ (Spain) through Grant No. PB97-1501 and from the Junta de
Extremadura (Fondo Social Europeo) through Grant No. PRI97C1041. M. L. H.
wants to thank the hospitality of Universidad de Extremadura, where the draft
of the paper was prepared.

\end{document}